\documentclass[journal=jpccck]{achemso}

\setkeys{acs}{etalmode=truncate}
\setkeys{acs}{maxauthors=10}
\setkeys{acs}{articletitle=true}
\usepackage{amssymb}

\usepackage{amssymb}

\usepackage{graphicx}
\usepackage{amsmath}
\usepackage{bm}
\usepackage{color, soul}
\usepackage[normalem]{ulem}
\usepackage{siunitx}

\usepackage[firstpage]{draftwatermark}
\usepackage[breaklinks]{hyperref}
\usepackage[all]{hypcap}

\title{Static and Dynamic Disorder in Triple-Cation Hybrid Perovskites}

\author{M.\ Baranowski}
\affiliation{Laboratoire National des Champs Magn\'etiques Intenses, UPR 3228, CNRS-UGA-UPS-INSA, Grenoble and Toulouse, France}
\alsoaffiliation{Department of Experimental Physics, Faculty of Fundamental Problems of Technology, Wroclaw University of Science and Technology, Wroclaw, Poland}
\author{J.\ M.\ Urban}
\affiliation{Laboratoire National des Champs Magn\'etiques Intenses, UPR 3228, CNRS-UGA-UPS-INSA, Grenoble and Toulouse, France}
\author{N.\ Zhang}
\affiliation{Laboratoire National des Champs Magn\'etiques Intenses, UPR 3228, CNRS-UGA-UPS-INSA, Grenoble and Toulouse, France}
\author{A.\ Surrente}
\affiliation{Laboratoire National des Champs Magn\'etiques Intenses, UPR 3228, CNRS-UGA-UPS-INSA, Grenoble and Toulouse, France}
\author{D. K.\ Maude}
\affiliation{Laboratoire National des Champs Magn\'etiques Intenses, UPR 3228, CNRS-UGA-UPS-INSA, Grenoble and Toulouse, France}
\author{Zahra Andaji-Garmaroudi}
\affiliation{Cavendish Laboratory, J.J. Thomson Avenue, Cambridge CB3 0HE, United Kingdom}
\author{S. D. Stranks}
\affiliation{Cavendish Laboratory, J.J. Thomson Avenue, Cambridge CB3 0HE, United Kingdom}
\author{P.\ Plochocka}\email{paulina.plochocka@lncmi.cnrs.fr}
\affiliation{Laboratoire National des Champs Magn\'etiques Intenses, UPR 3228, CNRS-UGA-UPS-INSA, Grenoble and Toulouse, France}

\begin{document}




\begin{abstract}
A detailed understanding of the carrier dynamics and emission characteristics of organic-inorganic lead halide perovskites is critical for their optoelectronic and energy harvesting applications. In this work, we reveal the impact of the crystal lattice disorder on the photo-generated electron-hole pairs through low-temperature photoluminescence measurements. We provide strong evidence that the intrinsic disorder forms a sub-bandgap tail density of states, which determines the emission properties at low temperature. The PL spectra indicate that the disorder evolves with increasing temperature, changing its character from static to dynamic. This change is accompanied by a rapid drop of the PL efficiency, originating from the increased mobility of excitons/polarons, which enables them to reach deep non-radiative recombination centers more easily.   

\end{abstract}

\maketitle

\section{INTRODUCTION}
Organic-inorganic perovskites have emerged as an extremely promising class of materials for energy harvesting applications. Perovskite based solar cells recently exceeded 20\% power conversion efficiency  \cite{saliba2016incorporation, yang2017iodide}, with an unprecedented fivefold increase in less than 10 years \cite{C8RA00384J}. This has led to a renewed interest in understanding their basic physical properties, which are responsible for the outstanding performance of these optoelectronic devices. The combination of a large optical absorption coefficient, long carrier lifetime and diffusion length in these materials  \cite{stranks2013electron,
vorpahl2015impact,hutter2017direct}, along with the simple and cheap fabrication with a minimal energetic footprint, makes organic-inorganic perovskites ideal candidates for both photovoltaic and light emitting applications 
\cite{yang2017recent,stranks2015metal}.

In organic-inorganic perovskites, the methylammonium (CH$_3$NH$_3$$^+$, MA) or formamidinium CH(NH$_2$)$_2$$^+$, FA) cation is surrounded
by PbX$_6$ octahedra. The band-edge states are composed essentially from the orbitals of the Pb and halide atoms  \cite{brivio2014relativistic, even2015pedestrian, yu2016effective}. The conduction band is composed of Pb p-type orbitals while the valance band is composed of Pb s-type orbital with admixture of p-type halide orbitals.  However organic cations also  play an important role in the electronic and optical properties of organic-inorganic perovskites \cite{dar2016origin, chen2016extended, chen2015rotational, wang2017indirect,
etienne2016dynamical, quarti2015structural, panzer2017impact, ma2017nature, motta2015revealing, ma2014nanoscale}. The motion of the organic cation and its local arrangement affect both the lattice parameters and the internal geometry of the crystal, lowering its symmetry\cite{motta2015revealing} at tetragonal and orthorhombic phase and locally breaking the symmetry on ps-time scale \cite{kubicki2017cation}. The wide range of possible bond angles between the organic cations and the halide cage gives rise to a plethora of distortions of the crystal lattice that can be simultaneously present in the perovskite structure \cite{motta2015revealing, ma2014nanoscale, quarti2015structural}. Detailed density functional theory calculations combined with molecular dynamics simulations reveal that these distortions affect the position of the bands and the local density of states. The resulting fluctuations in the potential landscape have been proposed to explain the modest carriers mobilities  \cite{ma2017nature} and can even change the nature of the fundamental band edge from direct to indirect
\cite{etienne2016dynamical, azarhoosh2016research, kepenekian2017rashba}. Moreover, a strong coupling between the lattice and free carriers/excitons can lead to local screening of carriers and the formation of polarons 
\cite{zhu2015charge,zhu2016screening,kang2017shallow, chen2016extended, neukirch2016polaron, miyata2017large, miyata2017lead, kubicki2017cation}. In the high temperature phase
($T>160$\,K for MA and $T>120$\,K for FA), the organic cations are free to rotate, and the induced disorder has a dynamic
nature \cite{poglitsch1987dynamic, kepenekian2017rashba, leguy2015dynamics, quarti2015structural, motta2015revealing,
ma2017nature}. At lower temperatures, the motion of organic cations becomes progressively more restricted
\cite{fabini2016dielectric, dar2016origin, chen2015rotational, leguy2015dynamics, mattoni2015methylammonium}. Nevertheless, the
temporary disordered configuration of MA or FA can be transferred to the low temperature orthorhombic phase, which leads to the
formation of distorted orthorhombic domains \cite{dar2016origin} or even a ``frozen'' tetragonal phase \cite{tahara2016experimental, galkowski2017spatially}. Thus, the presence of organic cations leads to an intrinsic disorder in the organic-inorganic lead halide perovskites, in addition to extrinsic disorder due to alloying or grain boundaries.

In this work, we use low-temperature photoluminescence (PL) and transmission spectroscopy to elucidate the basic photo-physical properties of the state-of-the-art triple cation  Cs$_{0.05}$(MA$_{0.17}$FA$_{0.83}$)$_{0.95}$Pb(I$_{0.83}$Br$_{0.17}$)$_{3}$ (TC) perovskite alloyed films \cite{soufiani2017impact}, for which a power conversion efficiency in
excess of 20\% has been reported \cite{saliba2016cesium}. The characteristic line shape of the PL indicates that the low temperature emission is dominated by carriers/excitons localized by shallow trapping states forming an exponentially vanishing tail density of states below the band edge. We find that the PL points to an evolution of the trapping state population with increasing temperature. Our results demonstrate a change in the character of the disorder from static to dynamic with increasing temperature. This is accompanied by a rapid drop of the PL intensity. In light of this, we discuss the fundamental impact of the below band gap states on the recombination process and their potential nature.

\section{METHODS}

Cover slips were cleaned by sonication in acetone and isopropyl alcohol for 30 min, then the substrates were further cleaned with an oxygen plasma treatment for 10 min. The Cs$_{0.05}$(MA$_{0.17}$FA$_{0.83}$)$_{0.95}$Pb(I$_{0.83}$Br$_{0.17}$)$_{3}$ perovskite precursor solution was prepared by dissolving PbI$_2$ (1.1 M), PbBr$_2$ (0.22 M), formamidinium iodide (1 M), and methylammonium bromide (0.20 M) in a mixture of anhydrous DMF:DMSO (4:1 volume ratio, v:v) followed by addition of 5 vol\% from CsI stock solution (1.5 M in DMSO). The perovskite solution was spin-coated on the glass substrates in a two-step program at 1000 and 4000 rpm for 10 and 30 s respectively, and 110 $\mu$l of chlorobenzene was poured on the spinning substrate 30 s after the starting of the program. The substrates were then annealed at \SI{100}{\celsius} for 1 hour. Synthesis and deposition of perovskite solutions were performed inside a nitrogen glove box under controlled moisture (5ppm) and oxygen conditions (0.5ppm). 

For optical spectroscopy, the samples were mounted in an exchange gas helium cryostat. Optical access to the samples was provided through a quartz
window. The PL measurements were performed in backscattering configuration. The sample was excited by a pulsed laser tuned to 540\,nm, with a pulse duration of
 300\,fs. The excitation beam was focused on the sample by a lens having a focal length of 20\,cm. The same lens was used to collect the PL
signal, which was analyzed using a 50\,cm focal length spectrometer equipped with a 600 grooves/mm grating and detected using liquid nitrogen cooled Si CCD camera. The transmission measurements were performed in the same setup, using the white light generated by a halogen lamp as broadband light source.
The signal for time resolved PL measurements was dispersed by a 30\,cm focal length monochromator with 300 grooves/mm grating, and detected using a streak camera.

\section{RESULTS \& DISCUSSION}

\begin{figure}[t!]
\centering
\includegraphics[width=0.6\linewidth]{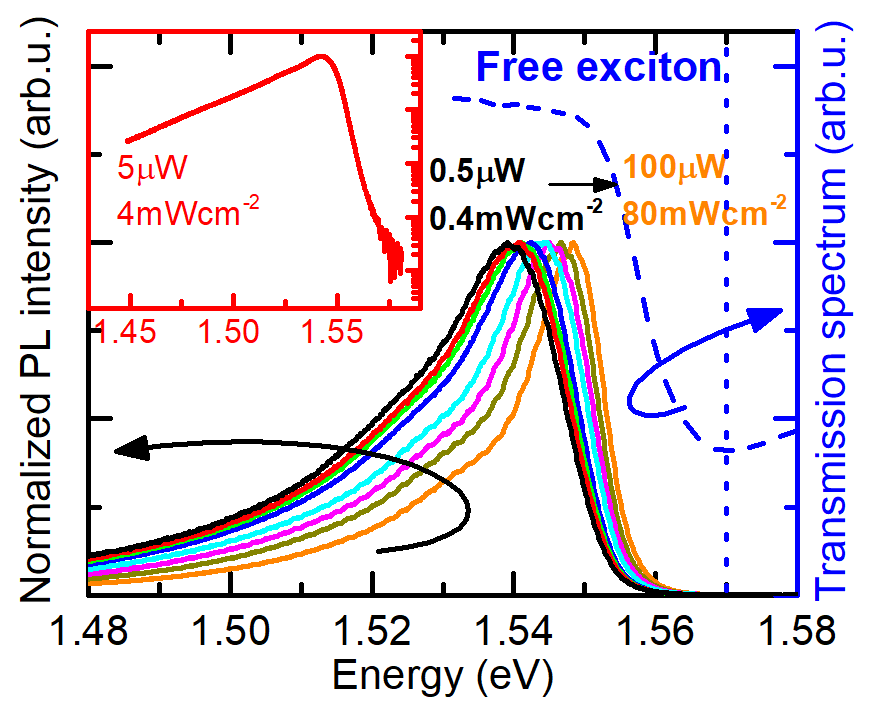}
\caption{Full lines: PL spectra of Cs$_{0.05}$MA$_{0.16}$FA$_{0.7}9$Pb(I$_{0.83}$Br$_{0.17}$)$_3$ thin films showing blue shift with increasing excitation power. Dashed blue line: transmission spectrum measured at the same position of the sample as PL spectrum. The PL spectra are measured with 0.5$\mu$W, 1$\mu$W, 2$\mu$W, 5$\mu$W, 10$\mu$W, 20$\mu$W, 50$\mu$W, 100$\mu$W excitation powers. The excitation wavelength is 540nm. Inset: PL spectrum plotted in logarithmic scale, revealing the low energy exponential tail. Measurements were performed at 1.5 K.} \label{fig:PL_low_T}
\end{figure}

Figure\,\ref{fig:PL_low_T} shows representative PL spectra measured at $T=1.5$\,K at different excitation powers, together with a transmission spectrum measured at the same position of the sample. The PL has a characteristic shape with a relatively sharp cut off on high energy side, and a long exponential tail on the low energy side, as shown more clearly in the semilogarithmic plot of the inset of Fig.\ \ref{fig:PL_low_T}. The PL peak is red shifted from the dip in the transmission spectrum, which corresponds to the free exciton (FX) absorption \cite{miyata2015direct, galkowski2016determination,yang2017unraveling,yang2017impact}. The magnitude of this red shift depends on the
excitation power, and gradually decreases with increasing power, while the PL line shape is
unaffected. This behavior is characteristic of PL associated with an exponential tail of density of states, as reported for inorganic semiconductors \cite{henini2004dilute, baranovskii1998temperature, chichibu2001localized, skolnick1986investigation}. In
semiconductors, the carrier/exciton relaxation rates are generally much higher than the recombination rates, so the photo-created
excitons thermalize (that is, are trapped to form localized excitons) before radiative recombination. In contrast, the absorption probes the
joint density of states, and therefore its resonance corresponds to the free exciton (FX) transition. The different time scale of relaxation and recombination explains the shift observed between the PL peak and the FX absorption energy as well as it power dependence. The continuous shift of the PL maximum indicates that the localized states form a
continuous distribution of energy levels -- a band tail \cite{wright2017band} -- rather than a single localized level. The exponential decrease of the PL spectrum at low energies shown in the inset of Fig.\ \ref{fig:PL_low_T} is a direct evidence for the exponential distribution with energy of the trapping states.

\begin{figure*}[t]
\centering
\includegraphics[width=0.95\linewidth]{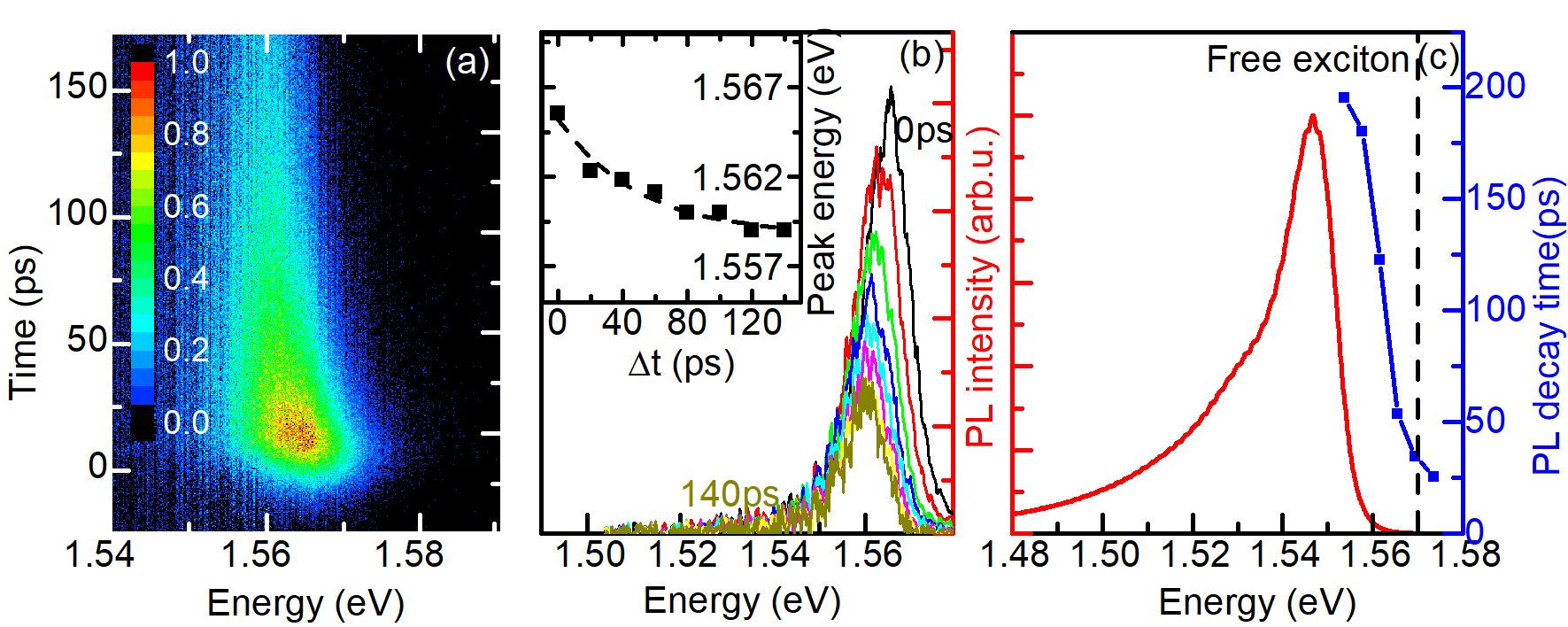}
\caption{(a) Streak camera image of initial 200 ps temporal dependence of the PL measured at $T$ = 1.5K with 50$\mu$W excitation power and 540nm excitation wavelength. (b) PL spectra at different times after excitation pulse (20 ps step). Inset: PL peak position as a function of time. (c) Time integrated PL spectrum measured with 50$\mu$W  excitation power (red curve) together with decay time constant energy dispersion (blue curve). The decay time is determined as the temporal distance between PL intensity maximum and the value where the PL intensity decreases by a factor of e.}\label{fig:TRPL}
\end{figure*}

The process of carrier relaxation through a tail density of states can be directly observed in time resolved PL measurements. Because the relaxation processes are much faster than recombination, this process is most prominent in the first few tens of ps of the PL temporal evolution.
Figure \ref{fig:TRPL}(a) shows a streak camera image of the temporal evolution of the PL, monitored up to 200\,ps after the excitation pulse, and measured at $T=1.5$\,K.
Characteristic features of PL dynamics in disordered systems \cite{buyanova1999mechanism, mair2000time} can be seen: (i) The PL
maximum shifts to lower energy with time (see Fig.\ \ref{fig:TRPL}(b)), and (ii) there is a strong energy dispersion of the PL decay as
function of energy (see Fig.\ \ref{fig:TRPL}(c)). The streak image very nicely illustrates the process of carrier relaxation through the
population of the localized states. Immediately after excitation, the PL peak is situated only a few meV below the energy of the free exciton, determined from the transmission spectrum (see Fig.\ \ref{fig:TRPL}(a,b)), and after red shifts continuously with time. The probability of free excitons being captured by trapping states with a given energy is proportional to the density of states. Therefore, just after excitation the PL spectrum reflects the shape of the tail of density states, and the PL peak is close to the free exciton energy (Fig.\ \ref{fig:TRPL} (b)). The later redshift of the PL maximum is induced by the thermalization of excitons to unoccupied, deeper states, i.e., driven by the tendency of the exciton population to reach thermal equilibrium with the lattice. Excitons trapped on shallow states act as a reservoir, which progressively populates the deeper states leading to a marked asymmetry of the PL decay time. It is important to note that the PL originating from donor-acceptor recombination can posses very similar characteristic features \cite{kong2015characterization}, i.e.\ strong Stokes shift dependence as a function of the excitation power and PL decay time asymmetry. However, we think that the strong dependence of the decay time at a given energy of detection on the excitation power (see Supporting Information) clearly points to the transfer process between localized states. Therefore, the application of the tail density of states picture is more accurate than the donor-acceptor model.




Importantly, our time resolved measurements performed at $T=1.5$\,K do not reveal any signature of hot photoluminescence previously
reported at room temperatures \cite{zhu2016screening, yang2017acoustic, bretschneider2017trap, fang2018long}. The motion of
organic cations at room temperature results in the ``dressing'' of the electrons and holes by the local organic sublattice
polarization. This protects carriers from efficient scattering by longitudinal optical (LO) phonons \cite{zhu2015charge, zhu2016screening}, elongating the
carrier relaxation to the band minimum and hot PL can be observed for few hundreds of ps up to a nanosecond after excitation
\cite{zhu2016screening, yang2017acoustic, bretschneider2017trap, fang2018long}. Our results demonstrate that at low temperatures, even very shortly after excitation, all the emitted photons have an energy below the free exciton transition. This indicates that the organic cation motion is frozen, which reduces the dynamic carrier screening\cite{zhu2016screening}. Therefore, LO phonon scattering can efficiently cool the photo-induced carriers down, in a similar fashion to III-V semiconductors\cite{shah2013ultrafast}.

\begin{figure}[h]
\centering
\includegraphics[width=0.95\linewidth]{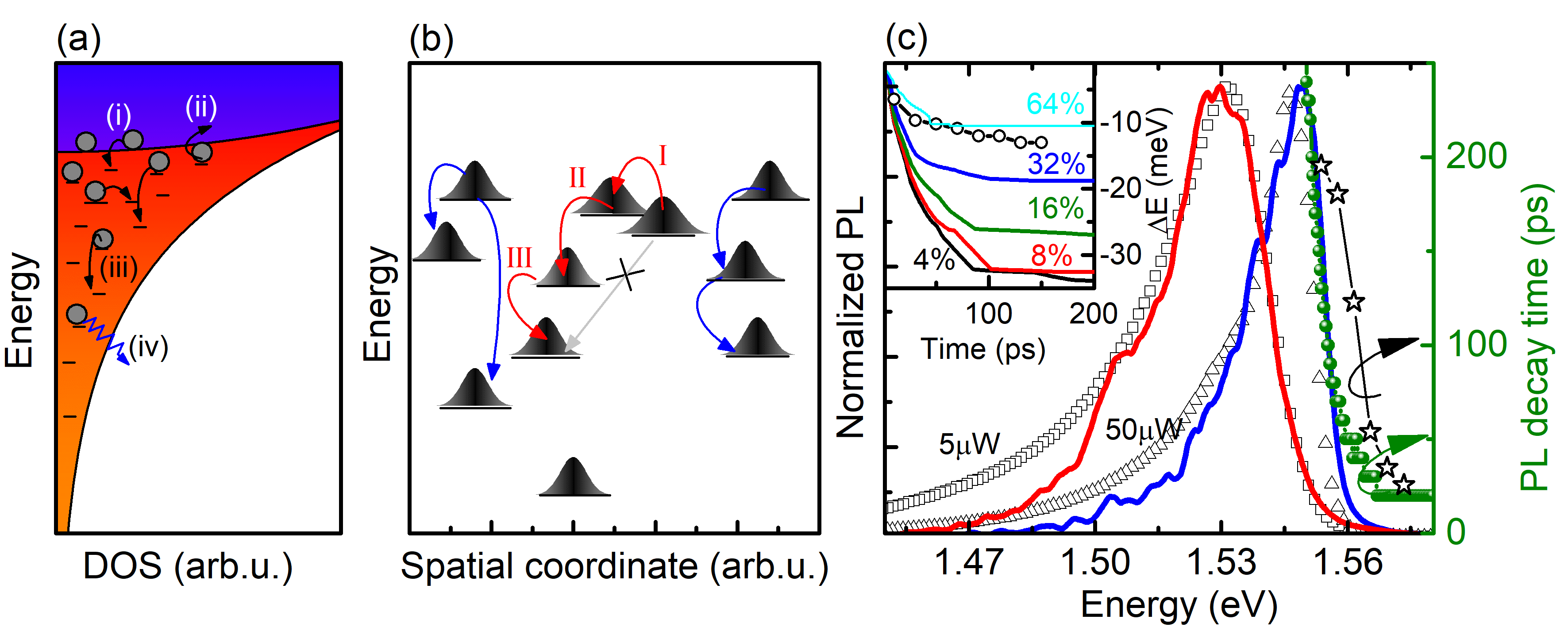}
\caption{(a) Scheme of possible processes taken into account in the simulation of exciton kinetics: (i) exciton trapping, (ii) exciton activation to band states, (iii) exciton hopping, (iv) exciton recombination. The orange part represents the energy distribution of trapping states and the blue part is a continuum of extended states (free exciton density of states). (b) Blue arrows indicate allowed carrier paths at very low temperature, i.e., carriers can only relax to states nearby and only with a lower energy. Red arrows indicate possible paths at slightly higher temperature. In this case, the carriers can also hop to near states with slightly higher energy (I). Therefore they became more “mobile” and can reach more global minimum in the process indicated by (II) and (III). This can lead to the observation of an initial redshift with slightly increasing temperature (see Fig.\ 4(c) between 1.5-\SI{25}{\K}). (c) Experimental PL spectra compared to predictions of the hopping model. The red and blue curves correspond to simulations performed with initial amount of injected carriers 8\% and 32\% of the number of trapping states to simulate different excitation powers. The squares and triangles are PL spectra measured at 5$\mu$W and 50$\mu$W excitation power. The green points correspond to simulated PL decay time dispersion. The stars are PL decay times determined experimentally. Inset: shift of the simulated PL peak position as a function of time for different trapping state saturation regimes together with experimental results (open points).} \label{fig:sym}
\end{figure}

Our PL data can be qualitatively explained using a hopping model previously employed for inorganic disordered semiconductors 
\cite{baranovskii1998temperature, rubel2006model, shakfa2015thermal, kazlauskas2005exciton, jandieri2013nonexponential,
baranowski2011model, baranowski2012dynamics, baranowski2015temperature}. The schematic of the model is presented in Fig. \ref{fig:sym}(a) and (b). The model assumes that excitons behave like
non-interacting single particles trapped by local potential minima with the possibility of excitons hopping between different
states, which have an exponentially decreasing density of states with decreasing energy. Non-radiative recombination is neglected, as we focus on the PL line shape rather than the intensity and the time scales of nonradiative recombination ($\sim$1-100 ns) are much slower than hopping and trapping processes ($\sim$1-100 ps). A trapped exciton can either recombine radiatively, hop to another localized state or eventually be activated to a free exciton state (mobility edge which provides the zero reference for energy in the model). The dynamics of excitons (PL) are determined by the rates of the competing processes. The hopping rate is described by the Miller-Abrahams formula  \cite{miller1960impurity, baranovskii1998temperature}:
\begin{equation*}
v_{ij}=v_{0}\exp\left (-\frac{2r_{ij}}{\alpha}\right )\exp\left ( \frac{E_i-E_j}{k_{\text{B}}T} \right ),
\end{equation*}
if $E_j>E_i$ and
\begin{equation*}
v_{ij}=v_{0}\exp\left (-\frac{2r_{ij}}{\alpha}\right ),
\end{equation*}
if $E_j \leq E_i$. Here, $v_{ij}$ is the hopping rate between states $i$ and $j$ with energy $E_i$ and $E_j$, $v_0$ is an
attempt rate, $\alpha$ is the decay length of the exciton wave function and $T$ indicates the temperature. The rate of exciton activation to the mobility edge (free exciton) is
\begin{equation*}
v_{\text{act}}=v_{0}\exp\left (-\frac{E_i}{k_{\text{B}}T}\right ),
\end{equation*}
and the radiative recombination rate is simply the inverse of exciton radiative life time $v_{\text{rec}}=\tau_{\text{r}}^{-1}$. The simulation
of exciton dynamics is performed on a randomly generated set of trapping sites (within a cubic volume of size $L^3$ with periodic
boundary conditions) with uniform position distribution and exponential energy distribution with a density of states
\begin{equation*}
D(E)=\frac{N}{L^3E_0}\exp\left (-\frac{E}{E_0}\right ),
\end{equation*}
where $E_0$ is the average localizing  energy and $N$ is the number of traps in a cube of size $L\times L \times L$. The
details of the approach used for the simulations can be found in the Supporting Information and in previous works
\cite{baranowski2013theoretical, baranowski2015nitrogen}. The carrier dynamics predicted by the model and the PL spectrum are
determined by the dimensionless parameters $\tau_{\text{r}} v_0$, and $\frac{N_0\alpha^3}{L^3}$, which can be treated as an effective density of states, and $E_0$. The average energy of localization can be estimated from the observed PL spectral broadening,
which depends weakly on other parameters \cite{rubel2005quantitative, baranowski2011hopping}. At very low temperature (${\text{k}}_{\text{B}}T\ll
E_0$), the full width half maximum (FWHM) of PL spectrum is FWHM $\sim 2.5E_0$. In our case, this leads to $E_0\approx 14$\,meV. We note that this value matches quite closely the experimentally measured values for the Urbach energy in related systems\cite{zhang2015enhanced, de2014organometallic}. The other parameters used in the simulation are $\tau_{\text{r}}v_0=2 \times 10^4$ and $\frac{N_0\alpha^3}{L^3}=0.088$
(see Supporting Information for more details). The PL intensity is assumed to be proportional to the number of occupied states at that energy at any given time.

The simulated PL spectra are presented in Fig.\,\ref{fig:sym}(c). For the chosen parameters, the shape of the simulated PL (solid lines) reproduces very nicely the measured PL spectra (symbols). The power-dependent shift of the PL peak as well as their temporal characteristics, such as decay time dispersion (green circle) and PL peak shift with time (inset), are all reproduced with a good qualitative agreement and reasonable quantitative agreement. Our model, combined with our experimental results, shows that at very low temperature the disorder and related shallow states present in organic-inorganic perovskites have a very similar impact on the PL as in case of “classic” inorganic semiconductors. The origin of potential fluctuation might be related to the point defects such as metal or halide vacancies, interstitials \cite{nan2018methylammonium, chen2018elucidating} or alloying effect. The redshift of PL spectra related to light induced halide migration can be probably excluded since it is not efficient at low temperatures\cite{hoke2015reversible} (see Supporting Information about stability of the investigated TC). The origin of the disorder can be also related to a different orientation of the frozen organic cations \cite{fabini2016dielectric,
dar2016origin, chen2015rotational, leguy2015dynamics, mattoni2015methylammonium} and tetragonal phase
inclusions \cite{tahara2016experimental, galkowski2017spatially}, which is supported by the temperature dependent PL measurements presented below.

\begin{figure*}[t]
\centering
\includegraphics[width=0.9\linewidth]{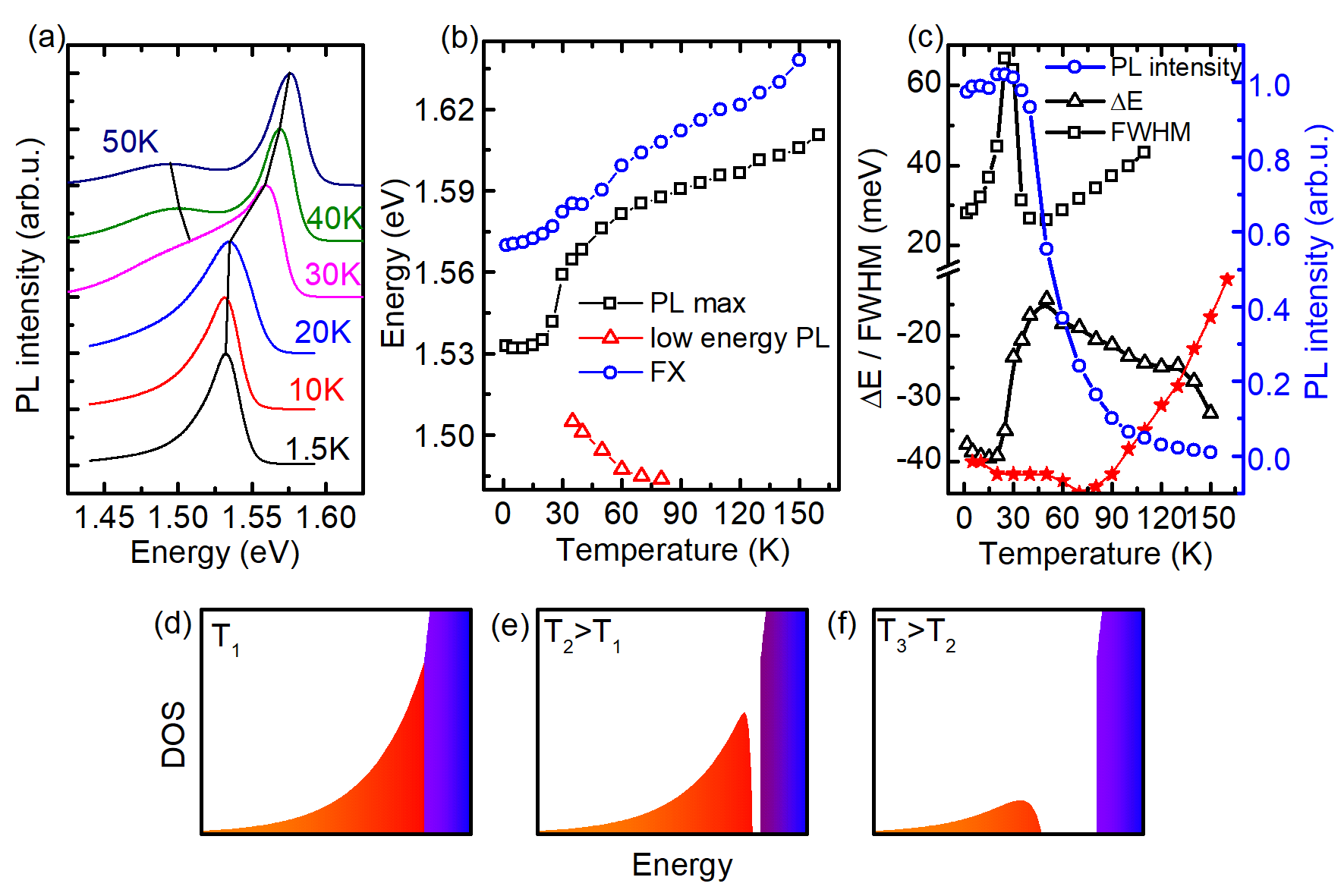}
\caption{(a) Time integrated PL spectra measured at different temperatures. The black lines are a guide to the eye. (b) Temperature dependence of free exciton transition (blue circle), main PL peak (black squares), and low energy PL peak (red triangles), (c) Stokes shift ($\Delta$E), PL broadening (FWHM), and PL intensity as a function of temperature. The red stars show results of Stokes shift simulation for an average energy of trapping states distribution of \SI{14}{\milli\eV}. (d)-(f) Schematics of a possible evolution of the density of states with increasing temperature, which explains the splitting observed in the PL spectra (orange: localized states. Blue: extended states).} \label{fig:temp}
\end{figure*}

Figure\,\ref{fig:temp}(a) shows the evolution of the PL spectra as a function of temperature. Initially, the main PL peak
position exhibits a small (few meV) red shift with increasing temperature. Above $T=20$\,K a second feature appears at lower
energies in the PL spectrum, while the main PL peak rapidly blue shifts (faster than the free exciton, determined via transmission measurements), which reduces the Stokes shift
(see Fig. \ref{fig:temp}(b)). Initially, the additional feature appears as a shoulder increasing the FWHM before two clearly
separated peaks can be observed above 30\,K (see Fig.\ \ref{fig:temp}(a,c)). The Stokes shift is $\sim 38$ \,meV at 1.5\,K, and it
decreases to around 15\,meV at $40-50$\,K (see Fig.\ \ref{fig:temp}(c)). In this so called ``S-shape'' behavior
\cite{kaschner2001recombination, cho1998s, baranovskii1998temperature,bergman2000photoluminescence, kopaczek2015optical}, the
initial red shift of PL peak is usually understood as a redistribution (optimization) of the distribution of the excitons, which acquire
enough thermal energy to leave their potential minima to find a lower energy state further afield. This behavior is correctly
reproduced by the hopping model (see Supporting Information). With a further increase of temperature, according to the Miller-Abrahams formula, the
motion of excitons between localized states becomes faster. This facilitates the establishment of a thermal distribution of the
excitons, which in turn induces a blue shift. The temperature for the onset of the blue shift depends on the average energy $E_0$ of
trapping states \cite{rubel2005quantitative} (see Supplementary Information) and it is given by $T \approx 0.55 E_0/k_{\text{B}}$.

Based on the estimated value of $E_0=14$\,meV from the low temperature (1.5\,K) measurements, the onset of the blue shift (decrease
of PL Stokes shift) should start at around 90\,K, which is a factor of three larger than the observed onset temperature of $\sim
30$\,K. In fact, such a significant blue shift between \SI{25}{\K} and \SI{50}{\K} is difficult to explain only based on carrier redistribution or activation to free states. This suggests that an additional mechanism is at work, which we attribute to microstructural evolution of the perovskite crystal with a transition from static disorder to dynamic disorder. According to theoretical predictions, different orientations of the frozen organic cations \cite{dar2016origin,chen2015rotational,leguy2015dynamics,fabini2016dielectric,mattoni2015methylammonium} and tetragonal phase inclusions \cite{tahara2016experimental,galkowski2017spatially} can produce potential landscape fluctuations giving rise to the observed tail density of states emission at very low temperature. In the low temperature orthorombic phase of MAPbI$_3$ or FAPbI$_3$, one would expect that the organic cations are all aligned in the same direction, because of the constraints imposed by the PbI$_6$ backbone. However, molecular dynamic calculations show that, upon cooling, the disordered configuration can also persist, which leads to a local bandgap modification and therefore to the formation of tail density of states \cite{dar2016origin}. This disordered orthorhombic domains are metastable states, thus, an increase of temperature leads to the transition to more ordered alignment of organic cations, i.e., formation of ordered orthorhombic phase and vanishing tail density of states. Additionally, the activation of the motion of the organic cations with increasing temperature can also screen the potential of point defects \cite{nan2018methylammonium}, which effectively can be treated as a reduction of their density.  The evolution of the density of trapping states is not only reflected by a rapid decrease of the Stokes shift, but also by the evolution of the PL line shape (Fig. \ref{fig:temp}(a)). The appearance of the second peak around \SI{30}{\K}, which progressively red shifts with increasing temperature, points to the melting of frozen domains from shallow to deeper ones, as schematically illustrated in Fig.\ \ref{fig:temp}(d-f). The temperature evolution of the PL is qualitatively very similar to that of  MAPbI$_3$\cite{dar2016origin}. However, the temperature at which the melting of the frozen domain begin is lower in the case of the triple cation perovskite (\SI{25}{\K}), as compared to \SI{50}{\K} for MAPbI$_3$. Conversely, this temperature is much closer to that of pure FAPbI$_3$\cite{wright2017band}, which is not surprising, considering the composition of the investigated TC perovskite. Also the low temperature FWHM of the PL spectrum of the TC perovskite film is similar to that reported for FAPbI$_3$ \cite{wright2017band, wright2016electron} (around \SI{30}{\milli\eV}), and significantly lower than in  MAPbI$_3$ \cite{wright2016electron,dar2016origin}, which can be attributed to the smaller dipole moment of formamidinium \cite{dar2016origin}.

The Stokes shift does not vanish completely above \SI{50}{\K}, which can indicate that the population of trapping states is not removed completely or that a new mechanism of excitons trapping is activated. Moreover, in contrast to III-V semiconductors \cite{kaschner2001recombination,bergman2000photoluminescence,
cho1998s}, the Stokes shift of the main PL peak begins to increase again at higher temperatures. The absolute value of the Stokes shift increases by a factor of two from $\simeq 15$\,meV at 50\,K to $\simeq 30$\,meV at 150\,K. This non-vanishing and increasing Stokes shift at high temperatures may be the  hallmark of the domination of dynamic disorder and  the formation of large polarons \cite{zhu2015charge,kang2017shallow, miyata2017lead, gong2016electron}: in case of fully static disorder, we would expect a progressive decrease of the Stokes shift. The transmission measurement provides information about the energy of the nascent, bare exciton transition, while the emission originates from the electron and holes dressed by the polarization of organic cations or inorganic lattice deformation. The exact nature of the polaron formation in the orthorhombic phase of perovskites is not obvious. The reorientation of the whole organic dipole is not possible. However, it has been demonstrated that the motion of the organic cations is not completely frozen, even at very low temperatures \cite{dru?bicki2016unexpected}, and possibly these remaining vibrations can play a role in the increasing Stokes shift observed above \SI{50}{\K}. The temperature at which this Stokes shift increases corresponds very well to the temperature of the glassy transformation observed in FAPbI$_3$ (FA which dominates over other cations in our triple cations sample)\cite{fabini2016dielectric}. Moreover, recent work shows that the polaron formation can be also related to inorganic lattice vibration/deformation \cite{park2018excited,miyata2017large,ambrosio2018origin}. For example, the energy of the I-–Pb–-I bending mode is 21cm$^{-1}$, therefore we can expect that it is active even at low temperatures. However, precisely understanding what type of organic or inorganic sublattice motion can lead to the formation of polaron states at low temperatures requires additional extensive theoretical studies. The increasing value of the Stokes shift with increasing temperature (in the high temperature range) might be attributed to the activation of new vibrational modes, which can result in more efficient coupling of carriers with the lattice and therefore increasing value of Stokes shift. 

The transition from a static disorder at very low temperature to a dynamic disorder at higher temperatures is accompanied by a
significant drop of the PL emission intensity (see blue points in Fig. \ref{fig:temp}(c)). This can be attributed to the increased mobility of excitons or polarons (large polarons behave like free carriers with an increased effective mass) when the population of trapping states and fluctuations are reduced with the initial increase of temperature (1.5-\SI{50}{\K}). It is well known that the PL efficiency and the concentration of non-radiative traps strongly varies from place to place in perovskite thin films \cite{vorpahl2015impact}. Therefore, we can expect that more
mobile excitons/polarons have an enhanced probability to reach non-radiative recombination centers than excitons localized on
traps induced by frozen disorder. A similar impact of the disorder on the PL efficiency has been reported in InGaN alloys \cite{nakamura1998roles}.

\section{CONCLUSIONS}

In conclusion, we have systematically investigated the exciton dynamics in TC thin films using PL and
transmission measurements. We show that at very low temperatures ($T < \SI{20}{\K}$) the PL possesses the characteristic features of the emission observed in inorganic disordered semiconductors. The power dependence of the Stokes shift, the energy dependence of the decay time and the low energy tail of the PL spectrum all point to a disorder induced exponential tail in the density of states. A hopping model for excitons qualitatively explains all the characteristic features of the PL spectra from our organic-inorganic lead halide perovskite thin films at very low temperature. The estimated average energy of the trapping states is \SI{14}{\milli\eV}. Unlike ``classic'' semiconductors, the temperature behavior of the PL indicates that the disorder and trapping state population evolve with increasing temperature. The character of disorder changes its nature from static to dynamic when lattice vibration couple with the carriers and probably lead to polaron formation. This is supported by a rapid decrease of Stokes shift and FWHM of PL, which cannot be explained by a simple carrier redistribution over the exponential tail in the density of states. The change of the disorder character is accompanied by a significant drop of the PL efficiency. We attribute this observation to the increased mobility of the excitons, which can reach more easily non-radiative recombination centers.

\section{ASSOCIATED CONTENT}
Details of hopping excitons model, power dependence of time resolved photoluminescence, XRD spectrum.

\begin{acknowledgement}
This work was partially supported by the R{\'e}gion Midi-Pyr{\'e}n{\'e}es under contract MESR 13053031, BLAPHENE and STRABOT projects, which received funding from the IDEX Toulouse, Emergence program,  ``Programme des Investissements d'Avenir'' under the
program ANR-11-IDEX-0002-02, reference ANR-10-LABX-0037-NEXT. N. Z.\ holds a fellowship from the Chinese Scholarship Council (CSC). M.B. appreciates support from the Polish Ministry of Science and Higher Education  within  the  Mobilnosc Plus program (grant no.\ 1648/MOB/V/2017/0). Z.A.-G. acknowledges funding from a Winton Studentship, and ICON Studentship from the Lloyd’s Register Foundation. S.D.S acknowledges the European Research Council (ERC) under the European Union’s Horizon 2020 research and innovation programme (HYPERION, grant agreement number 756962), and the Royal Society and Tata Group (UF150033). The work was supported by a Royal Society International Exchanges Cost Share award (IEC\textbackslash R2\textbackslash 170108).
\end{acknowledgement}

\bibliography{Bib}

\begin{tocentry}
\includegraphics{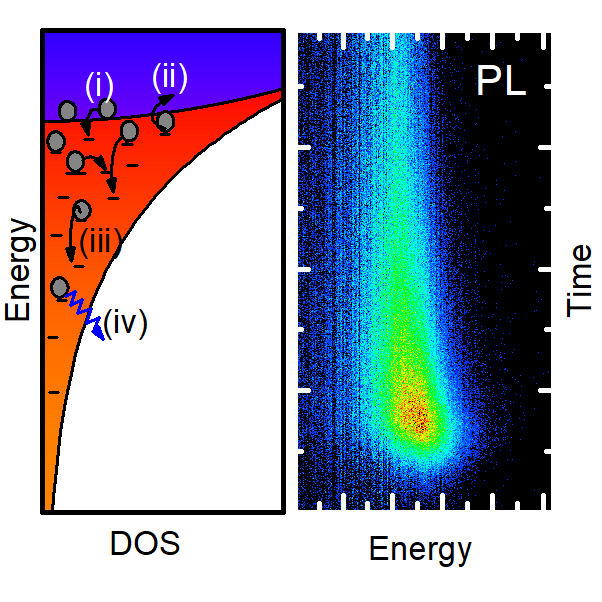}
\end{tocentry}


\end{document}